# Data-driven Co-clustering Model of Internet Usage in Large Mobile Societies

Saeed Moghaddam, Ahmed Helmy, Sanjay Ranka, Manas Somaiya

Computer and Information Science and Engineering (CISE) Department, University of Florida, Gainesville, FL

{saeed, helmy, ranka, mhs}@cise.ufl.edu

*Abstract* Design and simulation of future mobile networks will center around human interests and behavior. We propose a design paradigm for mobile networks driven by realistic models of users' on-line behavior, based on mining of billions of wireless-LAN records. We introduce a systematic method for large-scale multi-dimensional co-clustering of web activity for thousands of mobile users at 79 locations. We find surprisingly that users can be consistently modeled using ten clusters with disjoint profiles. Access patterns from multiple locations show differential user behavior. This is the first study to obtain such detailed results for mobile Internet usage.

## 1. Introduction

Wireless mobile networks are growing significantly in every aspect of our lives. Laptops, handhelds and smart phones are becoming ubiquitous providing (almost) continuous Internet access and ever-increasing demand and load on supporting networks. This provides new challenges and opportunities for the modeling and design of future mobile networks. By developing realistic behavioral models for mobile users' Internet access and website visitation patterns, novel behavior-aware network protocols can be developed and parameterized. Such behavioral models are essential and are established based on deep understanding of Internet usage of mobile users obtained through large-scale analysis of extensive wireless networks measurements. We refer to this approach as *data-driven* modeling and design paradigm.

The eventual goal of the data-driven paradigm is to utilize analysis of the users' behavior to drive the design of efficient context-aware protocols and services. This is in sharp contrast to the *general-purpose* design paradigm conventionally used in the wired Internet and much of the mobile (e.g., ad hoc) network design in the past decades. The general-purpose paradigm focuses on the design elements first, then evaluation using generic (usually random, uniform) models for traffic, Internet usage and mobility. Often, these models deviate dramatically from reality, which leads to sub-optimal performance or outright failure of the protocols during deployment. By contrast, the data-driven approach, shown in Fig. 1, starts by the analysis and realistic modeling of the target context and users, that then drives the design process.

This paper focuses on the fundamental first step in this paradigm; the data-driven realistic modeling of user Internet behavior. Particularly, we study models of correlated mobile user website access patterns, using clustering. Such behavioral clustering will aid in the understanding of the spatio-temporal load distribution on the network and similarity of users interest, and thus inform the design of important classes of applications, including modeling and scenario generation for network simulations, network capacity planning, web caching and interest-aware networking protocols [HDH08],[HDH09] to name a few.

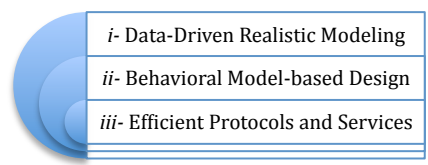

Fig. 1. The data-driven modeling and design paradigm

To obtain the behavioral model, we process extensive *netflow*, DHCP and MAC trap traces for thousands of mobile users in a WLAN spanning over 80 buildings and including over 700 APs, that we have collected. This represents by far the largest set of traces processed in any study of mobile networks to date. We provide a systematic method to process billions of records to integrate and aggregate the multi-dimensional data. The scale of the data provides a great challenge to employing advanced data mining techniques. We propose to use an information-theoretic co-clustering technique in a novel way to extract important relations between clusters of mobile users and clusters of accessed websites. We show that this method can provide accurate and efficient clustering with minimal information loss. The traces are then broken per building category, including educational, housing, health, cinema, etc. A location-based clustering is then carried out based on website visitation pattern similarity. Our method is systematic and can be generally applied to discover important spatio-temporal features of Internet behavior from other similar traces.

We report two major findings in this paper: 1- Mobile users cluster with respect to website visitation patterns into a small set of clusters that with clearly distinct profiles. For example, Mac users consistently visit washington post, cnet and apple websites and not microsoft. While PC users visit Yahoo!, Google and microsoft websites, but not apple's. 2- Locations in similar categories tend to cluster together, with a few exceptions, in terms of mobile website access. We establish the stability of these findings for three month-long samples. These findings provide the basis for mobile user behavioral models both qualitatively and quantitatively, as we discuss in our applications section.

Our work has the following key contributions:
1. We collect and process the largest set of mobile network usage traces (billions of records) and provide practical techniques for integrating and aggregating the data.
2. We propose an effective approach for multi-dimensional analysis of the dataset, and show how information theoretic co-clustering can be applied to create and correlate clusters of users and web domains to build group-specific profiles.
3. We conduct context-specific (location-aware) analysis of mobile users behaviors, using two different methods (hierarchical clustering and graph clique detection) to effectively discover groups of locations with similar contexts.
4. We obtain consistent results for clustering of mobile user and location behavioral similarity that provide the basis for future models of mobile Internet usage.

The rest of the paper is organized as follows. In Section 2, we review the related work. In Section 3, we address challenges associated with the collection, processing and analysis of large-scale wireless traces. Section 4 provides our case study using campus traces and co-clustering to develop realistic models of mobile societies. Section 5 discusses modeling and applications. Section 6 concludes.

## 2. Related Work

The rapid adoption of wireless communication technologies and devices has led to a widespread interest in analyzing the traces to understand user behavior. The scope of analysis includes WLAN usage and its evolution across time [TB00], [KE02], [HKA04], user mobility [HH06a], [BC03], [MV03], traffic flow statistics [MWYL04], user association patterns [PSS05] and encounter patterns [HH06b], [CHC+06]. Some previous works [HH06a],[HH06b] explore the space of understanding realistic user behaviors empirically from data traces. The two main trace libraries for the networking communities can be found in the archives at [ML09] and [CR09]. None of the available traces provides large-scale *netflow* information coupled with DHCP and WLAN sessions to be able to map IP addresses to MAC addresses to AP to location and eventually to a context (e.g., history department). *Therefore, (to the best of our knowledge) our work represents the first one to address large-scale multi-dimensional modeling of wireless and mobile societies. We analyze wireless mobile data around three orders of magnitude above any existing or ongoing study, providing finer granularity, richer semantics and potentially more accurate models. Our work also includes novel data processing techniques to address the challenges provided by this large-scale multi-dimensional data.*

There are several prominent examples of utilizing the data sets for context specific study. Mobility modeling is a fundamentally important issue, and several works focus on using the observed user behavior characteristics to design realistic and practical mobility models [HSPH07], [JLB05], [LKJB06], [KKK06]. They have shown that most widely used existing mobility models (mostly random mobility models, e.g., random walk, random waypoint; see [BH06] for a survey) fail to generate realistic mobility characteristics observed from the traces. Realistic mobility modeling is essential for protocol performance [BSH03a]. It has been shown that user mobility preference matrix representation leads to meaningful user clustering [HDH07]. Several other works with focus on classifying users based on their mobility periodicity [KK07], time-location information [EP06], [GBNQ06], or a combination of mobility statistics [TB04]. The work on the *TVC* model [HSPH07, HSPH09] provides a data-driven mobility model for protocol and service performance analysis. Our work is complementary to *TVC* and can extend *TVC* dramatically to incorporate dimensions of load, interest and web-site visitation preferences. In addition, the netflow traces are over three orders of magnitude more than the WLAN traces and the techniques used for their analysis and clustering are quite different. In [MWYL04] it was shown that the performance of resource scheduling [B03] and TCP vary widely between trace-driven analysis and non-trace-driven model analysis. *Using multi-dimensional modeling, our methods can develop new mobility-aware Internet-usage models, and utilize the realistic profiles to enhance the performance of networking protocols. Our new application of co-clustering techniques incorporates web activity, location and mobility, and provides user profiles users that may be used in a myriad of networking applications.*

One network application for multi-dimensional modeling is profile-based services. *Profile-cast* [HDH08],[HDH09], provides a new one-to-many communication paradigm targeted at a behavioral groups. In the profile-cast paradigm, profile-aware messages are sent to those who match a *behavioral profile.* Behavioral profiles in [HDH08,HDH09] use location visitation preference and are not aware of Internet activity. Other previous works also rely on movement patterns. *Our multi-dimensional modeling of mobile users, however, provides an enriched set of user attributes that relate to social behavior (e.g., interest, community as identified by web access, application, etc.) that has been largely ignored before.*

## 3. Modeling Approach

Data-driven modeling of large mobile societies requires three main phases to collect, process and analyze multi-dimensional large datasets with fine granularity (Fig. 2). In the first phase, extensive datasets are collected using the network infrastructure (or the mobile devices), plus augmenting information from online directories (e.g., buildings directory, maps) and the web services (e.g., whois lookup service). Data processing is the second phase to cross-correlate acquired information from different resources (e.g., access points, IP and MAC addresses), in which multiple datasets are manipulated, integrated and aggregated. The final phase is data analysis which includes global and location-based (context-specific) study of human behaviors based on their website access preferences and also the stability analysis of the findings for each case.

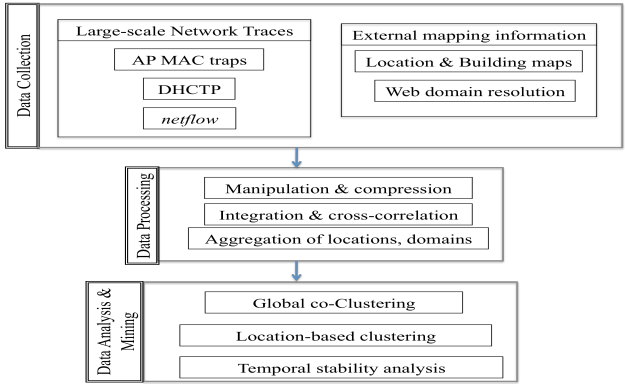

Fig. 2. Phases of modeling: collection, processing, and analysis.

## 3.1 Data Collection

For the campus-wide modeling of wireless users, we collect different types of traces via network switches including netflows, DHCP and wireless AP session logs (MAC traps). An IP flow is defined as a unidirectional sequence of packets with some common properties (e.g., IP address and port number source and destination) that pass through a network device (e.g., router). This device can be used for flow collection. The collected data provides fine-grained metering for detailed usage analysis. Network flows are highly granular; flow records include the start and finish times (or duration), source and destination IP addresses, port numbers, protocol numbers, and flow sizes (in packets and bytes) (see Table 1). The destination IP address can be used to identify the websites accessed, while the port and protocol numbers can identify the application used. The wireless session log is collected by each wireless access point (AP) or switch port (i.e., aggregate of APs in a building). The trace includes the 'start' and 'end' events for device associations (when they visited or left that specific AP), the device's MAC address, the date and time of those events, and the AP (or switch) IP and port numbers. From the above we can derive the association history (i.e., the location and time of user association) for all MAC addresses. The DHCP log contains the dynamic IP assignments to MAC addresses. The listed IP is given to the MAC address at the indicated date and time.

## 3.2 Data Processing

Data processing includes three steps of data manipulation, data integration and data aggregation to cross correlate the collected data before data analysis.

*Data Manipulation*

The variety and scale of different collected traces introduces one of the main challenges with respect to data manipulation. The size of the underlying data is very large and therefore, with a naïve approach the required time for each manipulating action would be in the order of a month, with tens of manipulations needed. For example, the netflow dataset gathered from USC campus includes around 2 billions of flow records for each month in 2008 which equals to 2.5 terabytes of data per year. Thus, appropriate methods for data manipulation are needed. Our approach to diminish the problem is to first compress the data via substituting similar patterns with binary codes and creating mapping headers to be used in future manipulation; then get the data exported into a database management system (MySQL) and finally design customized store procedures for the manipulation of data in a reasonable time.

*Data Integration*

The second requirement of multi-dimensional modeling is data integration. Data from different sources are not gathered in the same format and therefore a semantic link is required to be created in between them. For example, in our case study, users are represented by MAC addresses in wireless session logs and by IP addresses in netflow traces. However, when the data scale for one of the traces (in this case netflows) is very large, the cost of such integration using regular SQL commands increases dramatically. Thus, we designed customized stored procedures for this purpose.

*Data Aggregation*

Since the output of the integration process includes billions of records, we cannot directly feed the result to the analysis phase. Running rather any data mining method on such a large dataset will take years to accomplish. Therefore, we need an intermediate aggregation process for building design-specific views of the dataset. We can aggregate the records based on one or a set of fields e.g., time, user, location, domain name and application. The choice of appropriate aggregation scheme depends on the final design and modeling goals. If we are interested in studying usage patterns for different domains at different locations without considering single users or type of application, an aggregation on domain name, location and time for the number of bytes, packets or flows will be the best choice. If the goal is the study of users' spent time at different websites for different months, which is the case in our case study, we need to aggregate based on user id, domain name and months.

Table 1 – Netflow Sample

| Start Timestamp | Finish Timestamp | Source IP | Source Port | Dest IP | Dest Port | Protocol Num | ToS | Packet Count | Flow Size |
|---|---|---|---|---|---|---|---|---|---|
| 0618.00:00:07.184 | 0618.00:00:07.184 | 128.125.253.143 | 53 | 207.151.245.121 | 64209 | 17 | 0 | 1 | 469 |
| 0618.00:00:07.184 | 0618.00:00:07.472 | 207.151.241.60 | 52759 | 74.125.19.17 | 80 | 6 | 0 | 4 | 1789 |
| 0618.00:00:07.188 | 0618.00:00:07.188 | 193.19.82.9 | 31676 | 207.151.238.90 | 43798 | 17 | 0 | 1 | 103 |

### 3.3 Data Analysis

The data analysis phase is performed at three levels. The first level is to create a global model of dynamics within the network. This model is needed to provide a big picture of the dataset from the desired point of view. The next level is to build and analyze location and context-specific models, e.g., website access models in different types of buildings. The third level is to analyze the stability of learned models and get them revised if needed to get sufficient stability required for a solid model-based design.

*Global Analysis and Co-Clustering*

The main goal of global analysis is to provide a big picture of dynamics within the network. For this purpose, a very well known approach is to cluster entities (e.g., users, websites) with similar characteristics. However, a main challenge in modeling of multi-dimensional datasets is the fact that ordinary one-sided clustering algorithms like hierarchical clustering or k-means can only cluster data along different dimensions separately [JD88], i.e., either we get clusters of websites or clusters of users in our case. The proposed approach to resolve this problem is to apply co-clustering techniques, which cluster the input dataset along multiple dimensions simultaneously. In this way, we can correlate different dimensions in a unique model.

In our campus-wide case study, we first investigated applying bipartite graph co-clustering [D01]. A graph formulation is used in this algorithm coupled with a spectral heuristic (using eigenvectors) to co-cluster the 2-dimensional input data. However, a restriction of this algorithm was that each row cluster was associated with one column cluster, a restriction, which we found inappropriate to impose on our input dataset due to the variety of users' trends. Therefore, we instead choose another approach; the information theoretic co-clustering [DG03] for simultaneous clustering of users and domains to obtain a global model of the mobile society. We feed the users on-line activity matrix, which represents the time spent by users at different websites, into the algorithm. The theoretical formulation of this **co-clustering** technique treats the (normalized) non-negative contingency table as a joint-probability distribution of two discrete random variables, whose values are given in the rows and columns, and poses the co-clustering problem as an optimization problem in information theory. In this technique, co-clustering is performed by defining mappings from rows to row-clusters and from columns to column-clusters. These mappings produce clustered random variables. The optimal co-clustering is one that leads to maximum mutual information between the clustered random variables, and minimizes the *loss* in mutual information between the original random variables and the mutual information between the clustered random variables. This algorithm monotonically increases the preserved mutual information and optimizes the loss function by intertwining both row and column clustering. Row clustering is performed by calculating closeness of each row distribution (in relative entropy) to row cluster prototypes. Column clustering is performed similarly. This iterative process converges to a local minimum. This algorithm differs from one-sided clustering in that the row cluster prototypes incorporate column clustering information, and vice versa. The algorithm never increases the loss, and so, the quality of co-clustering improves gradually. It also ameliorates the problems of sparsity and high dimensionality. Iteratively, the method performs an adaptive dimensionality reduction and estimates fewer parameters than one-dimensional clustering approaches, resulting in a regularized clustering. In addition, the algorithm is efficient. The computational complexity of the algorithm is given by $O(N \cdot \tau \cdot (k + l))$ where $k$ and $l$ are the desired number of row and column clusters, $N$ is the number of non-zeros in the input joint distribution and $\tau$ is the number of iterations; empirically 20 iterations are shown to suffice.

The number of values along each dimension (represented by the number of mobile users and number of internet sites) is very large. Given, the size limitations of existing co-clustering algorithms and their implementation, we filter the dataset and limit ourselves to the most active websites and aggregate destination IP addresses and websites based on domains.

After executing the algorithm on the filtered dataset and getting the clusters of users and domains, we can create an association level matrix indicating the association level in between clusters along different dimensions. For each pair of a user cluster and a domain cluster, the association level is calculated by summing up the amount of all joint probabilities between them.

*Location-Based Analysis*

The main goal of location-based analysis is to discover different contextual clusters within a large mobile society. For this purpose, we first define a uniform schema for describing the context of a location. Then, we formulate a comparison method between different locations to find levels of context similarities. Finally, we devise an appropriate method to detect contextually similar locations.

As for the first step, the global acquired clusters of users and domains can be employed to create a uniform schema for context description. For each location, an association level matrix between groups of users and domains can be created using a uniform ordering of the acquired user and domain clusters. We employ the location specific association level matrix as a context descriptor of the location.

In the second step, we provide a method for comparing the context descriptors of different locations. For this purpose, we treat the corresponding association level matrices as vectors of all their values and employ cosine distance function. Using this method, we can create a dissimilarity matrix for different locations based on their context descriptors.

For the final step, we propose two different methods for finding groups of contextually similar locations. The first technique is to use hierarchical clustering to create clusters of locations. The second method is to map the dissimilarity matrix to a undirected graph as follows. Considering a node for each location, we draw an edge in between two different nodes if their dissimilarity is less than a threshold. Then, we find cliques within the graphs to discover groups of locations with similar contexts.

*Stability Analysis*

An important goal in data modeling is to discover stable models that can accurately describe not only the current state but also its time evolution and dynamics (i.e., its history and future). Such models are valuable in the sense that they can explain major trends during a long period of time (e.g., a semester) and thus can be effectively employed for realistic and durable model based designs.

To assess the relative stability of trends captured by global and location based models, we investigate whether the discovered clusters of users, domains and locations are sufficient to describe the history and the future of the mobile Internet access. Our method for measuring this forward, backward stability of the discovered clusters over time is to: 1- take all the interactions for the same sets of users, domains and locations during the previous and the next periods of the analyzed period; then, 2- for each of the periods, recreate the global association matrix and the location dissimilarity matrix using the same acquired clusters; and finally, 3- calculate the distance in between corresponding matrices for the analyzed period and the previous/next one. In this calculation, matrices are again treated as vectors of all their values and their distance is determined by cosine distance function.

## 4. Case Study and Experimental Results

In our case study, we conduct a campus-wide analysis on data we collected from the University of Southern California (USC) in 2008 based on the approach and techniques explain in the previous section.

### 4.1 Data Processing Details

The *netflow* and DHCP traces from the USC campus (over 700 access points) were processed to identify mobile user IDs using MAC addresses, and destinations, or 'peers' (usually web servers) using IP address prefixes. Over a billion records (for the month of March 2008) were considered initially, then the February and April traces (over two billions traces) were considered for the stability analysis. The IP prefixes (first 24 bits) were filtered using a threshold of 100,000 flows [the reason for using 24 bits filter is the fact that popular websites usually use an IP range instead of a single IP address]. For the filtered IP prefixes, their domains were resolved. Among the resolvable domains, the top 100 active ones were identified and all the users interacting with those domains (e.g., Google, Facebook, etc.) were considered for the analysis.

### 4.2 Global Analysis Results

A matrix was created associating the user IDs and domains (i.e., websites) using the corresponding total online time (per minute). For our analysis, we had 22,816 users, and 100 domains. The data is scaled using row-normalization of log the online time values. This is the input data for our modeling problem for which we applied the information theoretic co-clustering. In this case study, we discuss results for 'ten' clusters (i.e., with 'ten' as input to the algorithm for the number of output user and domain clusters[1]). Using this method of co-clustering, we produce two collections of domain clusters and user clusters, which are used to determine an association level between each pair of user and domain clusters. Fig. 3 shows the result of applying this method on the scaled data and Fig. 4 depicts the association level matrix between the resulting clusters. Each row in Fig. 4 identifies a group of users in terms of their association level to different groups of domains. In other words, each row provides a representation of users' interests to different groups of domains.

As shown, the algorithm is able to group users with similar access patterns into clusters. In a way, users within the same cluster maybe characterized by similar set of favorite wireless on-line activities. At a high level, we observe four general classes of user clusters with: (a) narrow access (cluster 1): users access two (or less) clusters of domains, in this case clusters *I* and *J* include usc and infoave (Telecom and webhosting), (b) narrow spread access (clusters 2,3): most user access time is spread over 3 or 4 domain clusters only, (c) medium spread access (clusters 4-7): most user access time is spread over 5 to 8 domain clusters, and (d) wide spread access (clusters 8-10): with noticeable user access in all domains. Cluster 7 depicts heavy Mac users. Again, as can be seen in the figure, these users rarely go to 'microsoft' but are interested in 'washingtonpost', 'cnet', Google, Yahoo!, Mozilla' in addition to social networking using 'facebook' and video sharing using 'youtube'. A deeper look into the clusters reveals some interesting trends. Clusters 2 and 3 include narrow spread users, but include clearly distinct sets of user interest. Cluster 2 shows users who mostly just utilize the Internet for search or email via Yahoo! and Google, and visit microsoft for probably getting software updates, and thus are likely Microsoft/PC users. Cluster 3, by contrast, shows users frequently go to 'apple' and 'mac' sites they rarely go to 'microsoft', and thus are likely mac users. Note that mac users are commonly interested in 'washingtonpost' or 'cnet'. Table 2 shows some other domains which are clustered together.

---

[1] Several values for the number of clusters (10, 20, 40) were investigated with 10 producing the best results.

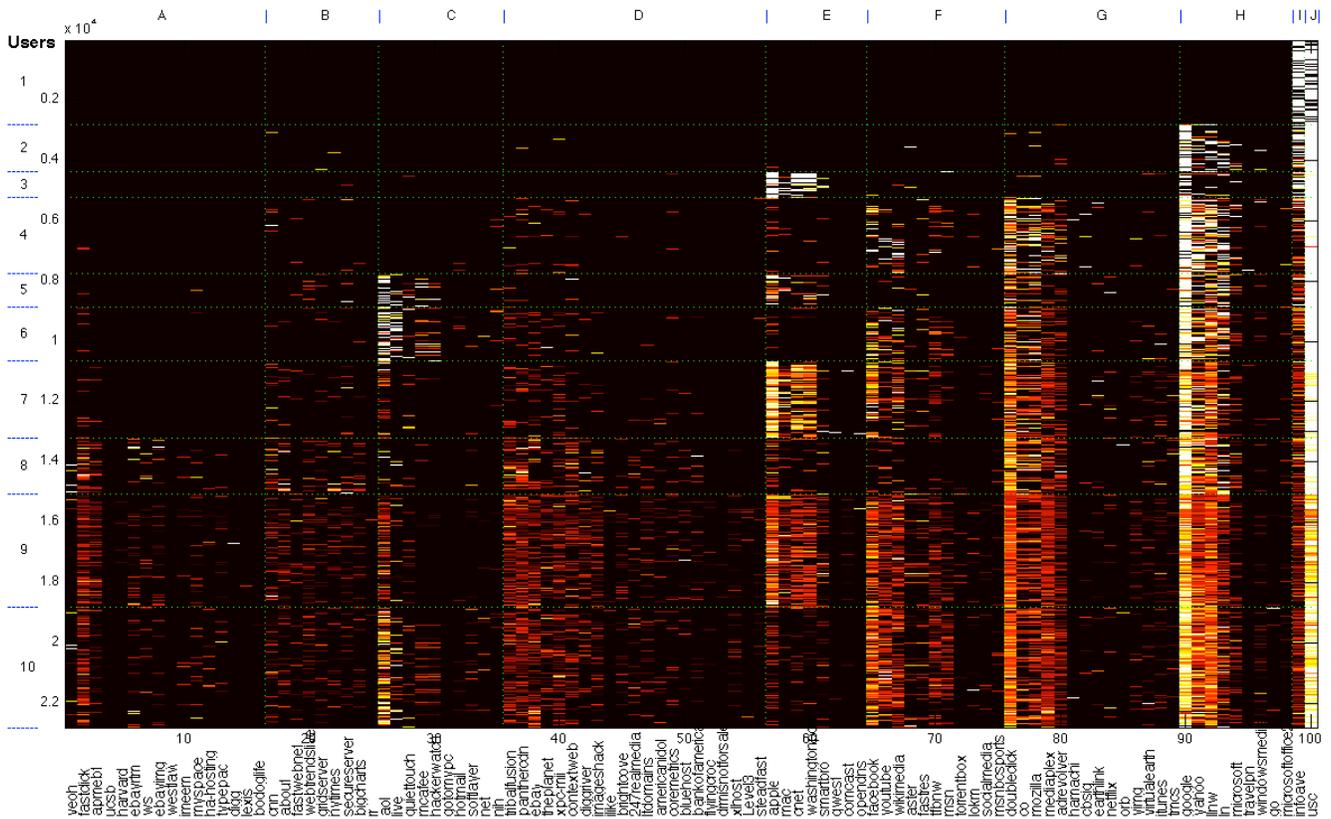

Fig. 3. Information theoretic co-clustering on user-domain matrix, March 2008. The result is given for ten clusters of users (1 through 10) and ten clusters of domains (*A* through *J*). Domain clusters *I* and *J* include one domain each. Colors in the range of black to white shows values in the range of minimum to maximum. X-axis shows the domain names and Y-axis indicates the user IDs.

Table 2 -Major related websites which are clustered together

| Cluster | Domains |
|---|---|
| A | myspace – imeem (social media service) - digg (social news) – typepad (blogging service)<br>ebayrtm - ebayimg - wsj (business news) - bodoglife(online gampling)<br>ucsb - harward - westlaw |
| B | cnn - new york times |
| C | mcafee - hackerwatch<br>live - hotmail |
| D | ebay - bankofamerica |
| E | apple – mac<br>washingtonpost - cnet |
| F | facebook – youtube - social media<br>msn - msnbcsports |
| G | netflix – itunes - orb (media cast)<br>tmcs (social city search) - virtualearth (online map) |
| H | google – yahoo<br>microsoft - windowsmedia - microsoftoffice2007 |

*Stability Analysis*: To assess the relative stability of trends, we process the records from Feb. 2008 and Apr. 2008 and recreate the global association matrices for them using the same clusters and ordering of users and domains in Fig. 3. The results are depicted in Fig. 5 and indicate that the trends hold to a large extent; the association level matrix for February is 92.25 percent and for April is 89.18 percent similar to that of March; plus, the association level matrices of February and April are 98.51 percent similar. This indeed indicates the stability of the results.

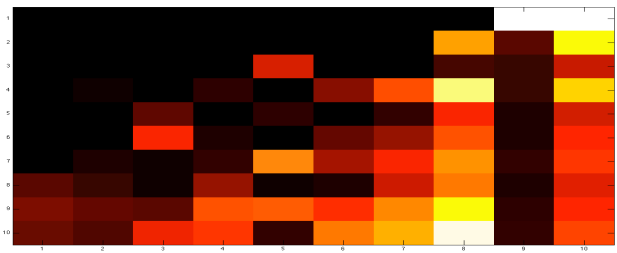

Fig. 4. Association level of resulting user and domain clusters by applying information theoretic co-clustering (March 2008)

### 4.3 Location-based Analysis Results

In the second phase, we model locations based on their acquired context descriptors and analyze the results based on the their actual context. For this purpose, for any interaction between a user and a domain, we first identify the switch-port that handled the connection using the

WLAN traces. Then, we associate the interaction to its location among 84 buildings across the campus using a mapping table between the switch-ports and buildings. Next, all active buildings (handling at least one interaction) in March 2008 are selected (79 buildings) and their context descriptors are created as explained in Section 3.3. Finally, we create dissimilarity matrix for all the selected buildings using the context descriptors and the metric explained in the same section. This matrix is used by two different techniques based on hierarchical clustering and graph clique detection to discover groups of contextually similar buildings. Fig. 6 shows the result of applying hierarchical clustering for creating 10 clusters of locations. In the figure, all clusters can be identified using green line borders and all distances (dissimilarities) can be figured using the z-axis. For each cluster, Fig. 7 shows the average dissimilarity between each corresponding building and all the others. To analyze the resulting clusters, we studied all the buildings and based on their actual context categorized them into 10 categories including: housing, auditorium, (outdoor) activity, sorority, fraternity, school, health, music, cinema and service. These categories are visualized by different colors in Fig 6.

As can be seen in Fig. 6, most of the buildings in the same category are clustered together into one or two clusters. For example sororities are all clustered together and fraternities form two major clusters and two uni-member clusters. The interesting point about the fraternities is the fact that those two uni-member clusters include professional fraternities and the other two contain social ones. We can also see that all auditoria are in the same cluster as well as cinema-related buildings. Regarding the "activity" category that includes buildings with different activity context including sports, religion, social and shopping, we notice that 6 out of 8 are in the same clusters while 3 of 6 are sports related. In addition, it can be observed that housing buildings form two major clusters and there is only one separated building in another cluster. The study of the building shows it is the only housing complex that includes a plaza and bookstore too. Health related buildings are also assigned into two main clusters. However, buildings in school and service categories are almost scattered among clusters because of the fact they include different types of schools for social work, journalism, humanities, letters and arts, law and leadership and different kind of centers for facilities management, financial, communication and computing services.

*Stability Analysis*: As before, to assess the relative stability of trends, we process the records from Feb. 2008 and Apr. 2008 and recreate the dissimilarity matrices for them using the same acquired clusters for March. The results are depicted in Fig. 8 and indicate that the trends hold to a large extent; the dissimilarity matrix for February is 92.72 percent and for April is 95.12 percent similar to that of March; plus, the dissimilarity matrices of February and April are 93.35 percent similar. This indeed indicates the stability of the results.

The second method detects cliques within the corresponding graph for dissimilarity matrix using the threshold of 0.06 as explained in Section 3.3. Fig. 9 shows the resulting graph layout for the data from March. As can be inferred from the histogram for the dissimilarity matrix (Fig. 10), the resulting graph includes around 10 percent of all possible edges using the mentioned threshold. As can be seen in the graph, a clear relationship exists between identified cliques and the actual categories of buildings.

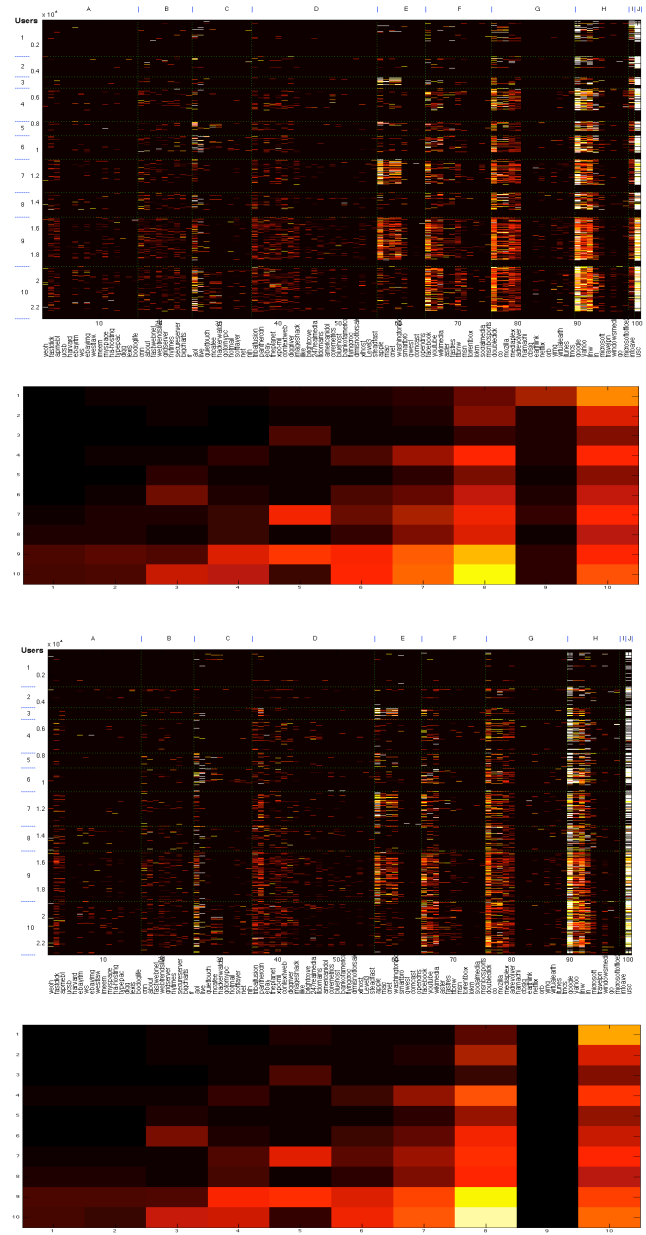

Fig. 5. Using the same column and row ordering of users and domains obtained via co-clustering of the March 2008 trace, these graphs are constructed using Feb 2008 (up) and Apr 2008 (down) measurements. The trends are relatively stable from one month to another especially for the narrow spread and wide spread clusters.

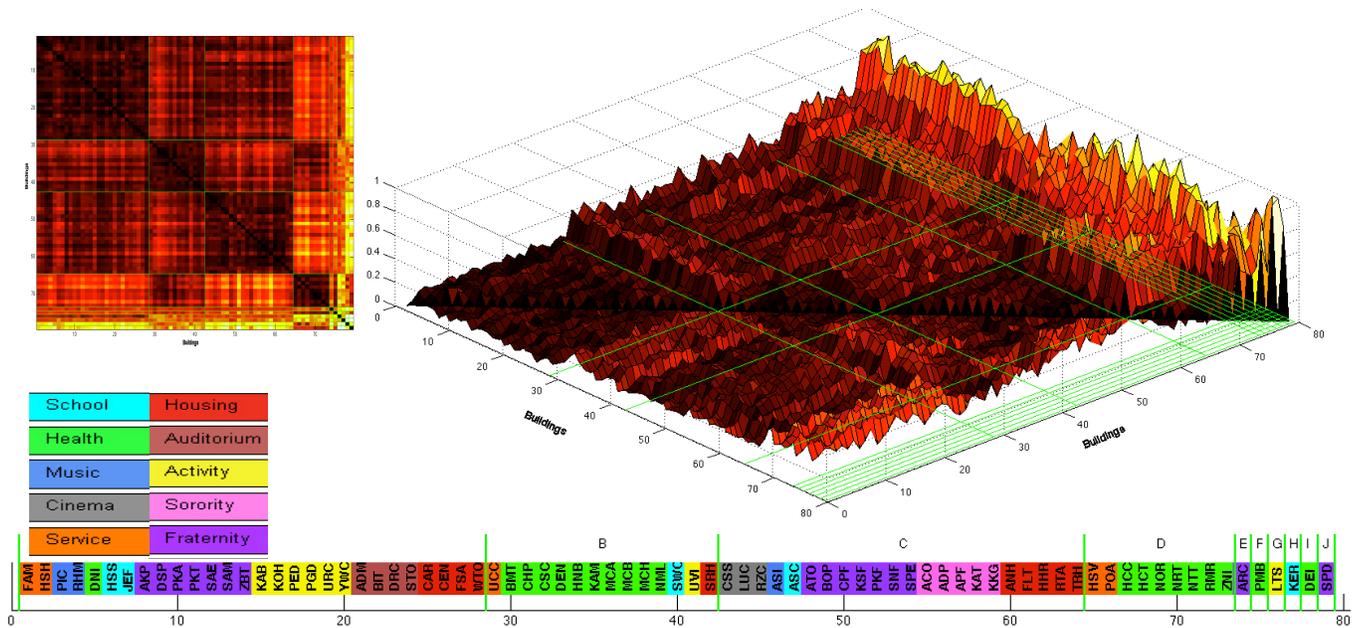

Fig. 6. Clusters of buildings for March 2008 distinguished by green lines; x and y axes show building abbreviations and z-axis shows their dissimilarities (distances) in between 0 and 1. The color of each abbreviation shows the context of the building based on the available color map in the picture.

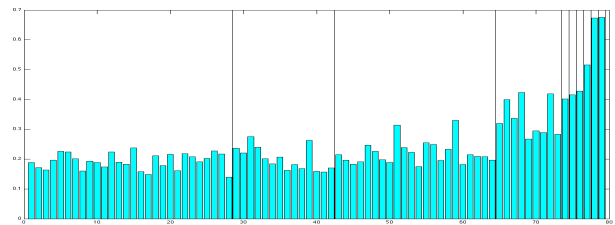

Fig. 7. Average dissimilarity between each building in a cluster and all the other buildings (March 2008)

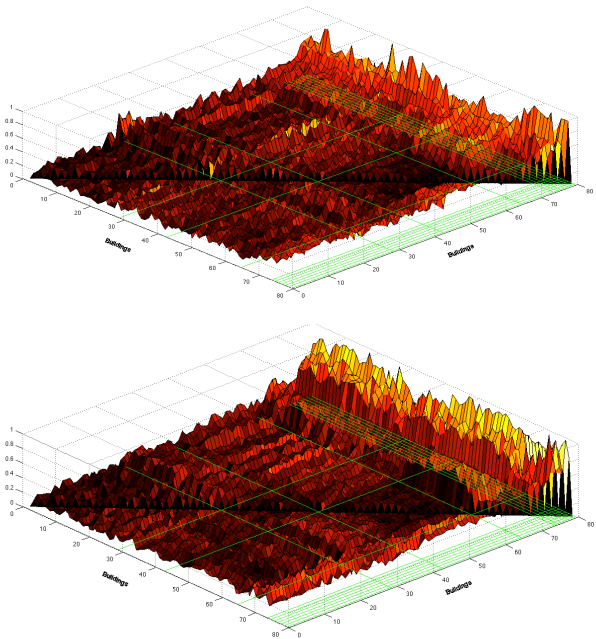

Fig. 8. Clusters of buildings for Feb. (up) and Apr. 2008 (down)

## 5. Discussion: Modeling and Applications

The systematic data-driven mining method proposed in this paper can be used with any set of wireless data to discover significant features that may be used as a similarity metric for mobile users. The method, and our findings from the global and location-based analyses can be used in several important applications in mobile networking research. Here, we specifically address (albeit briefly for lack of space) three such major applications:

1- Modeling and simulating spatio-temporal web usage for mobile users: Network simulations are essential for the design and evaluation of mobile networks (e.g., ns-2). To provide realistic input to the simulations, realistic models of node behavior are needed, along with scenarios of events and dynamics of mobility, traffic and Internet access. While earlier work has focused on mobility and traffic modeling, we provide the first work on modeling of mobile Internet website clustering. The spatio-temporal parameters of on-line activity (in terms of time of access, duration and location context), along with correlation and clustering between nodes in the simulation can be easily derived from our analysis in this paper. None of the existing models captures such spatio-temporal correlation across website access, nor does it capture correlations between nodes in that aspect. Recreating network usage more accurately will result in significantly different node density, load, and similarity distributions from those created by today's models.

One model that can benefit directly from our analysis is a mobile Internet usage (website access) model, with inputs including number of user-website clusters, distribution of user and website cluster sizes along with access pattern characterization (i.e., wide or narrow spread). Similarly, for location categories and clusters of buildings, the spatial distribution of website access patterns can borrow from our findings in the location-based analysis.

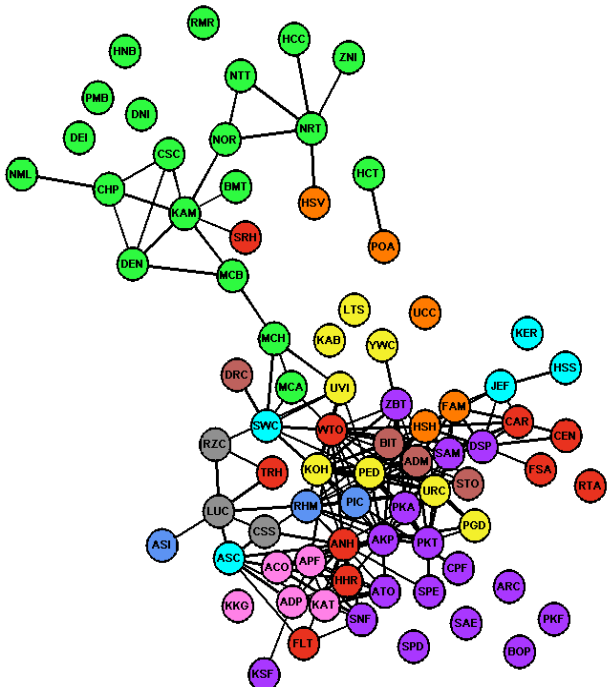

Fig. 9. Graph representation of dissimilarity matrix using the threshold of 0.06 for March 2008

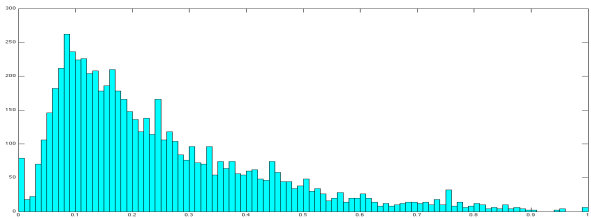

Fig. 10. Histogram for the dissimilarity matrix for March 2008

Developing and releasing the code for the mobile Internet access model is part of our future work. Similarly, we plan to conduct an extensive study on the spatio-temporal parameters for mobile traffic modeling in the future.

2- Interest-based protocols and services: A new class of protocols and services center around user-interest and similarity, including profile-cast, trust establishment [KTH10], participatory sensing [RES10], crowd sourcing, location-based services, alert notification and targeted announcements and ads. So far, mobility patterns (e.g., in profile-cast) have been used to infer interest. Website access patterns can significantly enhance the accuracy of interest inference and provide much needed granularity for these protocols and services. The interest models developed based on our analysis can aid both the informed design of such efficient protocols and the realistic evaluation thereof.

3- Network planning and web caching: Load distribution on the network is essential for network capacity planning and on-going configuration and management issues, and is certainly related to web access patterns and its clustering. Also, the caching of web objects for mobile users can only be efficient if informed by the history of access patterns. These applications for mobile networks are becoming more compelling especially with the significant increase of usage of smart phones, iphones, ipads, and the like.

## 6. Conclusion

This study is motivated by the need for a paradigm shift that is data-driven to develop realistic models and efficient protocols for the future mobile Internet. We provided a systematic method to process and analyze the largest mobile trace to date, with billions of records of Internet usage from a campus network, including thousands of users and dozens of buildings. Novel analysis was conducted utilizing advanced data mining using efficient co-clustering, at the global and location-based levels. We have shown that mobile Internet usage can be modeled with a strikingly small number of clusters of distinct web access profiles. Similarly, building categories show very distinct Internet usage patterns and are often clustered together. These trends were found to be highly stable over time.

The details of our study enable the parameterization of new and realistic models for mobile Internet usage with applications in several areas of networking, including simulation and evaluation of protocols, mobile web caching, network planning and interest-aware services, to name a few. We hope for our method and analysis to provide an example for large-scale data-driven modeling of mobile networks in the future. With more measurements from mobile and sensor networks becoming available, we expect our method to facilitate analysis of many other large datasets in future studies.